\documentclass[3p,times]{elsarticle} 

\usepackage{amssymb}
\usepackage{amsmath} 
\usepackage{amsxtra}
\usepackage{mathrsfs}
\usepackage{bm}  
\usepackage{subfig}
 
\usepackage{xcolor}

\journal{Journal}

\begin{document}

\begin{frontmatter} 

\title{Electro-diffusive modeling and the role of spine geometry on action potential propagation in neurons}
 
\author[add1]{Rahul Gulati}
\author[add1]{Shiva Rudraraju}
\ead{shiva.rudraraju@wisc.edu} 
\address[add1]{Department of Mechanical Engineering, University of Wisconsin-Madison, WI, USA}

\begin{abstract}
Electrical signaling in the brain plays a vital role to our existence but at the same time, the fundamental mechanism of this propagation is undeciphered. Notable advancements have been made in the numerical modeling supplementing the related experimental findings. Cable theory based models provided a significant breakthrough in understanding the mechanism of electrical propagation in the neuronal axons. Cable theory, however, fails for thin geometries such as a spine or a dendrite of a neuron, amongst its other limitations. Recently, the spatiotemporal propagation has been precisely modeled using the Poisson-Nernst-Planck (PNP) electro-diffusive theory in the neuronal axons as well as the dendritic spines respectively. Patch clamp and voltage imaging experiments have extensively aided the study of action potential propagation exclusively for the neuronal axons but not the dendritic spines because of the challenges linked with their thin geometry. In the last few years, assisted by the super-resolution microscopes and the voltage dyeing experiments, it has become possible to precisely measure the voltage in the dendritic spines. This has facilitated the requirement of a high fidelity numerical frame that is capable of acting as a digital twin. Here, for the first time, using the PNP theory, we integrate the dendritic spine, soma and the axon region to numerically model the propagation of excitatory synaptic potential in a complete neuronal geometry with the synaptic input at the spines, potential initiating at the axon hillock and propagating through the neuronal axon. The model outputs the scintillating forward propagation of the action potential along the neuronal axons as well as the back propagation into the dendritic spines. We point out the significance of the intricate geometry of the dendritic spines, namely the spine neck length and radius,  and the ion channel density in the axon hillock to the action potential initiation and propagation. We also study the role of excitatory synaptic input in multiple dendritic spines to the neuronal communication. 
\end{abstract}


\end{frontmatter}

\section{Introduction}
\label{section: introduction}

The brain, a complex structure, has the neuron as its fundamental building block. The primary role of the neuron is to smoothly transmit the electrical signals also known as the action potential. The understanding of the propagation of this action potential is critical to understand the functioning of the brain, the disruption of which can lead to neurological disorders and a number of diseases such as, Alzheimers, traumatic brain injury etc \cite{marion2018tbi, palop2007alzheimers, palop2010alzheimers, ghatak2019mechanisms, chaudhury2015depression}. The primary mechanism of the inter-neuronal electrical transmission is through the release of neurotransmitter by the pre-synaptic neuron which activates the receptors on the post-synaptic neuron. The excitatory synaptic potential is thus initiated at the dendritic spines, accumulated at the axon hillock until the threshold is attained leading to the firing of the all-or-none action potential along the axon. Besides the role of the different components of the neuron, such as the spine, the dendritic shaft, the axon hillock and the axon, the transmission is aided by the rich ion channel density and the receptor density present on the neuronal membrane.

The neuron consists of thousands of dendritic spines, which are the narrow protrusions on the dendritic shaft. The spines were discovered more than a century ago and their importance as the principal site for excitatory synaptic input was established. The neurotransmitter receptors present on the spine are activated once the neurotransmitter binds onto them leading to the influx of excitatory input in the form of synaptic current. It has also been contested that the geometry of the spine plays a paramount role in regulating the synaptic potential and therefore in synaptic plasticity. Experimental investigation of the spines has been challenging due to difficulty in accessing them owing to their small size. Numerical modeling in the form of cable theory has been widely applied to understand the fundamentals of the electrical propagation but it was realized that this theory breaks down for thin compartments as the spine \cite{qian1999pnp}. The Poisson and Nernst-Planck equations consider the coupling of the electric field to the ionic flow and provide a high-fidelity capability to model the rich spatio-temporal propagation of the action potential. Recently, dendritic spines have been numerically modeled using the PNP electrodiffusive equations. Coarse grained PNP equations are modeled by \cite{lagache2019compneuroscience}, PNP equations under electroneutrality assumption is considered by \cite{jerome2018physical},  PNP equations are modeled by \cite{cartailler2018neuron}. \cite{cartailler2018neuron, lagache2019compneuroscience} emphasize the role of the spine geometry on potential induced at the soma by the synaptic input at the spine. It is also to be noted that their numerical models do not include the ion channels on the spine as they do not play a prominent role during excitatory synaptic input \cite{palmer2009neuroscience}. 

Recently, super-resolution imaging techniques and fluorescent voltage dyes have enabled the visualization of the potential in the spines \cite{holcman2015nature}. The effect of attenuation of the excitatory synaptic input from the spine to the soma has been studied by \cite{acker2016eneuro, kwon2017cell, palmer2009neuroscience, jayant2017nature, yuste2013neuroscience, harnett2012nature, cartailler2018neuron}.  Spines with longer neck or lesser radius are electrically silent at the soma. \cite{harnett2012nature, araya2006pnas}  have predicted biochemical compartmentalization at the spine heads due to the high diffusional resistance offered by the spine neck. Effect of backpropagating action potential in spines has been closely observed by \cite{jayant2017nature, acker2016eneuro, kwon2017cell, palmer2009neuroscience}. It has been inferred that the back-propagation leads to high potential in the spines and the role of the ion channels in the spines is not as prominent during the excitatory synaptic input. There is a high density of potassium ion channels in the dendrites which ensures that the action potential does not initiate in the dendrite \cite{hoffman1997nature}. It has also been observed that there is a high density of sodium ion channels in the axon hillock that results in attaining the threshold potential in the hillock region \cite{colbert1996neuroscience}. With the rising capability of the experimental investigation of the spine biophysics, it has become obligatory for the numerical models to faithfully supplement them and act as their digital twin. 

Once the voltage threshold is attained at the axon hillock, it leads to the propagation of action potential in the axon. Neuronal axon electrophysiology has been studied through varied experimental investigations such as patch-clamp technique, electroencephalograms (EEG), electrocardiogram (ECG), MRI, voltage imaging etc which have consequently provided immense details of the voltage profile, conduction speed etc. Hodgkin-Huxley based cable theory provided a first mathematical basis to model the electrical transmission in the axon \cite{hodgkin1952huxley, huxley1959resistance}. With advanced imaging techniques, the presence of myelin sheath and the peri-axonal space have lead to the single-cable and double-cable models \cite{cohen2020cell}. The cable theory based models have a number of limitations and this lead to the electro-diffusive PNP modeling \cite{qian1999pnp, pods2013pnp}. Both the unmyelinated and the myelinated case have been modeled for a single node of ranvier \cite{lopreore2008pnp2, dione2016pnp1}. A first complete model of propagation of action potential along the myelinated neuronal axon consisting of multiple nodes of Ranvier using PNP was demonstrated by \cite{gulati2023brain} which has immense capability to act as a digital twin for the electrophysiological experiments. 

Currently, the numerical models of action potential propagation using PNP are limited to a spine only or the axon only propagation. With the advancements of the experimental investigations into the spine, there is a requirement of an integrated numerical model that demonstrates propagation of the electrical signaling starting from the spine and propagating to the axon through the dendritic shaft and the soma region. This unified model can aid the experimental findings and provide insights to a far greater resolution. In this work, we extend our previous work \cite{gulati2023brain} to include the spine, dendrite and the soma region to model the action potential propagation in the complete neuronal geometry. To our knowledge, this is the first integrated model which is based on the Poisson-Nernst-Planck equations. It is to be noted that there are full neuronal scale numerical models that exist in literature but they are based on the cable theory which is inaccurate for small compartments as the dendritic spine \cite{popovic2015nature, acker2016eneuro, harnett2012nature}. We demonstrate the initiation and propagation of action potential in the neuron, namely, the forward propagation along the axon and the back propagation into the spines. The effect of spine geometry on voltage attenuation at the soma is also studied. It is believed that spines modify their geometry and play an  essential role in synaptic plasticity. This model can also be coupled with mechanics to represent the complex neuronal mechano-electrophysiological interactions \cite{deb2021shellMechanics}.

\section{Methods}

\subsection{A field theoretic model of action potential propagation}
\label{section: PNP}

Cable theory based electrical network models have proven to be fundamental to our current understanding of the dynamics involved in action potential propagation. However, at the core, the network models are reduced order representations that try to capture the complex spatio-temporal variations of the ionic transport, voltage distribution, and most importantly the membrane structural heterogeneity into effective electrical properties like capacitance and resistance of the membrane and the channels. A deeper investigation into spatial and temporal interactions of the ionic transport with the membrane microstructure (ion channel and pump distributions, myelin, glial environment, etc.) is desired, but the resulting coupled evolution of the voltage distributions becomes necessary when more complex phenomena like neuronal injury effects on action potential propagation \cite{bar2016strain, li2011effects, estrada2021neural}, and conditions like neuronal hyperexcitability observed with Alzheimer's disease \cite{palop2016network, ghatak2019mechanisms, kim2007bace1} are to be modeled. \\
An electro-diffusive framework for modeling spatio-temporal ionic charge distribution and the resulting voltage evolution using high-fidelity partial differential equations (PDE)  modeling coupled electrostatics and electrochemistry can be capable of faithfully representing spatio-temporally heterogenous evolution of the ionic and voltage distributions leading to generation, propagation and potentially disruption of the neuronal action potential. Such a capability to model spatial heterogeneity and membrane geometry-action potential interactions is evidently more important in the dendrites than the axons, due to the complex morphology of neuronal dendritic structures and synapses. {We now propose a PDE based field theoretic implementation of the Poisson-Nerst-Planck (PNP) framework that can model 2D/3D ionic and voltage field distributions and their interactions with the membrane microstructure.} \\
In this model, the transport of the respective ionic species, $c_i$, due to the corresponding diffusive and electromigration flux, $\boldsymbol{F}_i$, is modeled using the classical Nernst-Planck equation:

\begin{equation}
     \frac{\partial c_i}{\partial t} = - \nabla \cdot \boldsymbol{F}_i \label{eq:PNPConc}
\end{equation}

The ionic flux comprises of the diffusion term and the electro-migration term as presented in Eq. \ref{eq:PNPFlux}. Here $D_i$ is the diffusion coefficient of the i\textsuperscript{th} ion, $R$ is the gas constant, $T$ is the temperature, $F$ is the faraday constant and $z_i$ is the valency of the i\textsuperscript{th} ionic species.  It is to be noted that the cable theory based models only account for the electro-migration term i.e. the voltage gradient term. While the diffusion term may be neglected for axons with larger diameter, for finer geometries like dendrites, the diffusion term takes over the electro-migration term \cite{qian1999pnp}. Hence, the PNP treatment has a higher-fidelity in representing the underlying action potential propagation. The total flux is given by:

\begin{equation}
    \boldsymbol{F}_i = -[D_i (\nabla c_i + \frac{c_i F z_i}{RT} \nabla V )] \label{eq:PNPFlux}
\end{equation}

The coupling between the voltage and local ionic distribution is modeled with Poisson's equation \cite{pnp2008gauss2pnp}, shown below.
\begin{equation}
    -\nabla \cdot (\epsilon \nabla V) = F \sum_{i=1}z_i c_i \label{eq:poisson}
\end{equation}
where $\epsilon$ is the permittivity of the medium. In this work, we consider three ionic species: Na\textsuperscript{+}, K\textsuperscript{+} and Cl\textsuperscript{-}. The presence of an anion species ensures regulation of the required potential - for instance at the resting state. The sodium and potassium ions, as is well understood, depolarise and re-polarise the neuron. While the above PDE formulation by itself is well known, the novelty of its implementation in this work is from its application to model field variations and interactions for a complete neuron geometry considering the spine, dendritic shaft, soma, axon in the presence of geometric heterogeneity of the nodes of Ranvier, myelin distribution, and the ion channel density.

\subsection{Numerical implementation of the PNP model}
\label{mathematical formulation}
The coupled nonlinear system of PDE's for ionic concentration and electrical potential in the electro-diffusive PNP model are solved using the standard Finite Element Method (FEM). The primal fields that are solved for are the voltage and the concentration of Na\textsuperscript{+}/ K\textsuperscript{+}/ Cl\textsuperscript{-} ions. The salient features of the computational implementation are: adaptive mesh refinement near the nodes of Ranvier, adaptive time-stepping schemes, support for parallel direct and iterative (Krylov-subspace) solvers with Jacobi/SOR preconditioning. The weak formulation of the governing equations solved with FEM are given in the following subsection. The computational framework is made available to the wider research community as an open source library~\cite{GitRepo2022}, and we hope it serves as a platform for wider adoption of the high-fidelity PNP framework by neuronal modeling researchers.

\subsubsection{Weak formulation of the PNP model}
\label{sec:feModelling}
The Nernst-Planck and Poisson equations, expressed in their weak (integral) formulation that is suitable for the FEM implementation, following standard notation, are as follows:

\vspace{0.2in} 

\noindent Find the primal fields $\{ V, c_{Na}, c_{K}, c_{Cl} \}$, where,

\begin{align*}
V &\in \mathscr{S}_{V},  \quad \mathscr{S}_{V} = \{ V  ~\vert V   = ~\bar{V} ~\forall ~\textbf{X} \in \Gamma^{V}_g \}, \\
c_{i} &\in \mathscr{S}_{c_i},  \quad \mathscr{S}_{c_i} = \{ c_i  ~\vert c_i   = ~\bar{c}_i ~\forall ~\textbf{X} \in \Gamma^{c_i}_g \}
\end{align*}

and $i \in \{Na^+, K^+, Cl^-\}$, such that,   
\begin{align*}
\forall ~w^{V} &\in \mathscr{V}_{V},  \quad \mathscr{V}_{V} = \{ V  ~\vert V   = ~0 ~\forall ~\textbf{X} \in \Gamma^{V}_g \}, \\
\forall  ~w^{c_i} &\in \mathscr{V}_{c_i},  \quad \mathscr{V}_{c_i} = \{ c_i  ~\vert c_i   = ~0 ~\forall ~\textbf{X} \in \Gamma^{c_i}_g \}
\end{align*}
we have,
\begin{equation}
    -\frac{F}{\epsilon} \int_{\Omega} w^V~(c_{Na}+c_{K}-c_{Cl}) ~dV  +\int_{\Omega} \nabla w^V  \cdot \nabla V ~dV - \int_{\Gamma^{V}_h} w^V~(\nabla V \cdot \boldsymbol{n}) ~dS =0 \label{eq:weakPoisson}
\end{equation}
and,
\begin{equation}
    \int_{\Omega} w^{c_i}~\frac{\partial c_{i}}{\partial t} ~dV  + \int_{\Omega} \nabla w^{c_i} \cdot ~D_i \nabla c_i ~dV + \int_{\Omega} \nabla w^{c_i} \cdot ~D_i \frac{c_i z_i F}{R T} \nabla V ~dV  + \int_{\Gamma^{V}_h} w^{c_i} ~(\boldsymbol{F}_i \cdot \boldsymbol{n}) ~dS = 0 \label{eq:weakNP}
\end{equation}
where $w^{V}$ is the variation for the voltage field, $w^{c_i}$ are the variations for the ionic fields. $\Omega$ is the problem geometry, $\{ \Gamma^{V}_g, \Gamma^{c_i}_g \}$ are the Dirichlet boundaries and $\{ \Gamma^{V}_h, \Gamma^{c_i}_h \}$ are the Neumann (Flux) boundaries of the voltage and ionic fields, respectively. $\boldsymbol{n}$ is the unit normal vector . In this work, there is no voltage flux ($\nabla V \cdot \boldsymbol{n} =0$) at all the boundaries. Eq. \ref{eq:weakNP} and Eq. \ref{eq:weakPoisson} are the governing equations solved using FEM. \\

The PNP framework models the evolution of the voltage field and the ionic concentrations of Na\textsuperscript{+}, K\textsuperscript{+} and Cl\textsuperscript{-}. The initial ionic concentrations in the various regions such as the extra-cellular region, membrane, myelin, cytoskeleton, etc., are mentioned in the Supplemental Information. The initial voltage in the ECM is taken to be 0 mV and in the cytoskeleton region to be the resting value of -70 mV. The boundary conditions on the top and bottom surface of the domain are applied so that the fields, i.e. voltage and the ionic concentrations, represent their bulk value in the extra-cellular region as in Eq. \ref{eq:dbc}. The ionic exchange at the nodes of Ranvier is incorporated as an ionic flux as given by Eq. \ref{eq:nbc}. Here $I_i$ is the current of each ionic concentration computed using their respective Hodgkin-Huxley ionic conductance.\\

At top and bottom boundaries;
\begin{equation}
    c_i = c_i^e, \qquad V=0 \label{eq:dbc}
\end{equation} 

At the Node of Ranvier:
\begin{equation}
    \boldsymbol{F}_i \cdot \boldsymbol{n} = f_i, \qquad f_i = \frac{I_i}{z_i F} \label{eq:nbc}
\end{equation}

where, 
\begin{equation}
    I_{Na} = \bar{G}_{Na} m^3 h (V_m-E_{Na}) \nonumber
\end{equation}
\begin{equation}
    I_{K} = \bar{G}_K n^4 (V_m-E_K) \nonumber
\end{equation}
\begin{equation}
    I_{Cl} = 0 \nonumber
\end{equation}

The following expression for the leak current in the dendrite is adopted. 
\begin{equation}
    I_K^{leak} = g_K^{leak} (V-E_K)
\end{equation}

\subsubsection{Model parameters}

\begin{figure}[!h]
\centering
\includegraphics[width=1.0\linewidth]{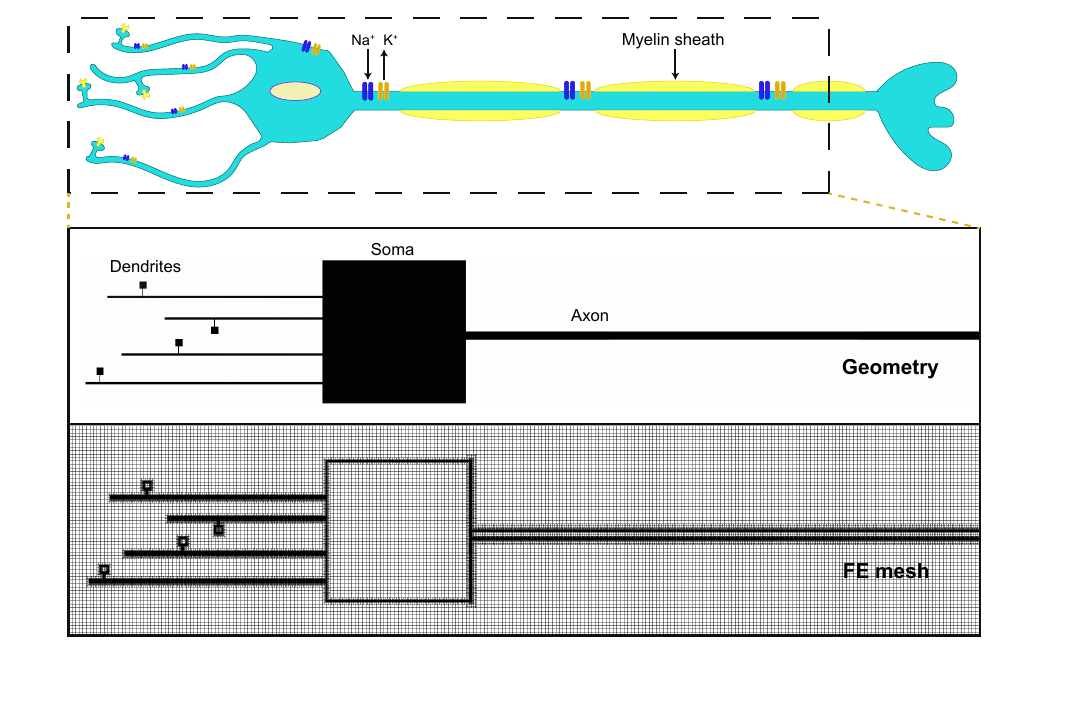}
\caption{{\bf The schematics of a simplified neuron geometry with respect to the electrophysiological modeling.}
The receptors in the spine head form the primary site of the synaptic input which leads the potential to flow through the spine neck, the dendritic shaft, the soma to the axon hillock where it accumulates until the threshold is attained and leading to the firing of this action potential along the axon. The myelin sheath adds to the faster propagation of this potential via spatio-temporal saltatory conduction. The geometry and the mesh used to represent the neuronal geometry for the finite element method are also presented.}
\label{fig1}
\end{figure}

Initially, the membrane potential is at its resting value of $-70$ mV with higher concentration of sodium ions in the extracellular region and a higher concentration of potassium ions in the intra-cellular region. The model is first equilibrated for a few timesteps until there are no fluctuations in the field variables. Next, to initiate the action potential, we assume that once the neurotransmitter released by the pre-synaptic neuron is captured by the receptors residing on the spine head of the post-synaptic neuron, it leads to sodium ion influx. This leads to the voltage buildup at the axon hillock region leading to the firing of the action potential once the threshold potential is attained. There are Hodgkin-Huxley based Na\textsuperscript{+}, K\textsuperscript{+} ion channels on the dendritic shaft, the soma and the axon neuronal membrane. Besides these, some leak K\textsuperscript{+} ion channels are also present on the dendrite which ensure that the potential threshold is not attained in the dendrite region. There is a high density of ion channel density in the axon hillock region which ensures the propagation of the action potential. The myelin sheath leads to the presence of nodes of Ranvier region on the axon which consist of ion channels. Fig. \ref{fig1} depicts the schematics of the neuron with a simplified geometry under investigation using the FE method. The geometrical, electrical and ionic parameters of the model are listed in Table \ref{table1}, \ref{table2} and \ref{table3}.

\begin{table}[!ht]
\centering
\caption{
{\bf Constants used in the electro-diffusive numerical framework.}}
\begin{tabular}{c c c}
\hline
\multicolumn{1}{|l|}{\bf Parameter} & \multicolumn{1}{|l|}{\bf Value} & \multicolumn{1}{|l|}{\bf Description} \\ [0.5ex]
R &	8.31454 $J \cdot mole^{-1} \cdot K^{-1}$ &	Gas constant \\  [0.5ex] \hline 
T &	279.45 $K$ &	Temperature \\ [0.5ex] \hline 
F &	96485 $C \cdot mole^{-1}$ & Faraday constant \\ [0.5ex] \hline
 $D_{Na}$ & 1.33 $\mu m^2 \cdot ms^{-1}$ & Diffusion coefficient of sodium ion \\ [0.5ex] \hline
$D_{K}$ &	1.96 $\mu m^2 \cdot ms^{-1}$ & Diffusion coefficient of potassium ion \\ [0.5ex] \hline
$D_{Cl}$ & 2.0 $\mu m^2 \cdot ms^{-1}$ & Diffusion coefficient of chloride ion \\ [0.5ex] \hline
$\epsilon_o$ & 8.88541 $e^{-12} C \cdot m^{-1} \cdot V^{-1}$ & Electric permittivity in vacuum \\ [0.5ex] \hline
$\epsilon_w$ & 80.0 & Relative dielectric permittivity of water \\ [0.5ex] \hline
$g_{Na}$ & 120 $mS/cm^2$ & peak conductance of $Na^+$	\\ [0.5ex] \hline
$g_{K}$ &	36 $mS/cm^2$ & peak conductance of $K^+$	\\ [0.5ex] \hline
$g_{Na}^{leak}$ & 0.05 $mS/cm^2$ & leak conductance of $Na^+$	\\ [0.5ex] \hline
$g_{K}^{leak}$ & 0.5 $mS/cm^2$ & leak conductance of $K^+$	\\ [0.5ex] \hline
$c_{Na}^i$ & 10 $mM$ &	Initial intracellular concentration of $Na^+$	\\ [0.5ex] \hline
$c_{K}^i$ & 140 $mM$ &	Initial intracellular concentration of $K^+$	\\ [0.5ex] \hline
$c_{Cl}^i$ & slightly exceeds 150 $mM$ & Initial intracellular concentration of $Cl^-$	\\ [0.5ex] \hline
$c_{Na}^e$ & 155 $mM$ & Initial extracellular concentration of $Na^+$	\\ [0.5ex] \hline
$c_{K}^e$ & 3.5 $mM$ & Initial extracellular concentration of $K^+$	\\ [0.5ex] \hline
$c_{Cl}^e$ & 158.5 $mM$ & Initial extracellular concentration of $Cl^-$	\\ [0.5ex] \hline
\end{tabular}

\label{table1}

\end{table}

\begin{table}[!ht]
\centering
\caption{
{\bf Geometric properties of the neuron studied for the action potential propagation using the PNP model.}}
\begin{tabular}{c c}
\hline
\multicolumn{1}{|l|}{\bf Parameter} & \multicolumn{1}{|l|}{\bf Value} \\ [0.5ex]
Spine head dimension & 1.0 $\mu m$*1.0 $\mu m$  \\ [0.5ex] \hline
Spine neck length & 1.0 $\mu m$  \\ [0.5ex] \hline
Spine neck diameter & 0.1 $\mu m$  \\ [0.5ex] \hline
Dendritic shaft diameter & 0.3 $\mu m$  \\ [0.5ex] \hline
Distance of Spine 1 from soma & 25 $\mu m$ \\ [0.5ex] \hline
Distance of Spine 2 from soma & 15 $\mu m$ \\ [0.5ex] \hline
Distance of Spine 3 from soma & 20 $\mu m$ \\ [0.5ex] \hline
Distance of Spine 4 from soma & 31 $\mu m$ \\ [0.5ex] \hline
Soma dimension & 20 $\mu m$*20 $\mu m$  \\ [0.5ex] \hline
Ion channel length at dendritic shaft & 1.0 $\mu m$  \\ [0.5ex] \hline
Ion channel length at soma & 3.0 $\mu m$  \\ [0.5ex] \hline
Node of Ranvier length in axon & 5.0 $\mu m$  \\ [0.5ex] \hline
Ion channel length at the axon hillock & 50.0 $\mu m$  \\ [0.5ex] \hline
Axonal internodal distance & 70 $\mu m$  \\ [0.5ex] \hline
Diameter of rat axon & 1.1 $\mu m$  \\ [0.5ex] \hline
\end{tabular}
\label{table2}
\end{table}

\begin{table}[!ht]
\centering
\caption{
{\bf Electrical properties of the neuron under investigation.}}
\begin{tabular}{c c c}
\hline
\multicolumn{1}{|l|}{\bf Parameter} & \multicolumn{1}{|l|}{\bf Value} & \multicolumn{1}{|l|}{\bf Description} \\ [0.5ex]
$C_{Spine}$ / $C_{Dendrite}$ / $C_{Soma}$  & 0.85 $\mu F/cm^2$ & Membrane capacitance \\ [0.5ex] \hline
$C_{Axon}^{Myelin}$    & 0.16 $\mu F/cm^2$ & Myelin sheath capacitance \\ [0.5ex] \hline
$C_{Axon}^{Membrane}$   & 1.45 $\mu F/cm^2$ & Membrane capacitance \\ [0.5ex] \hline
$R_{Neck}$ & 54 $M \Omega$ & Spine neck resistance \\ [0.5ex] \hline
\end{tabular}
\label{table3}
\end{table}

\section{Results}

\label{section: results}
This section seeks to establish the significance of the PNP numerical framework by analyzing the results of the initiation and forward / backward propagation of action potential in the neuronal geometry. First and foremost, the synaptic current is input at a single spine head leading to the initiation of the axon potential at the axon hillock and propagating along the axon. Subsequently, the effect of geometry of the spine and the dendritic shaft are carefully studied and detailed insights are presented. Finally, we consider the case of synaptic input at multiple spine heads and present an exhaustive comparison with the corresponding case of input at a single spine head. Neuron of a rat with myelinated axon is considered for the numerical simulations. The parameters of the axon are collected from \cite{hodgkin1952huxley, cohen2020cell, dione2016pnp1, lopreore2008pnp2, gulati2023brain}.

\subsection{Synaptic input at single spine head}
\begin{figure}[!h]
 \centering
 \includegraphics[width=0.5\linewidth]{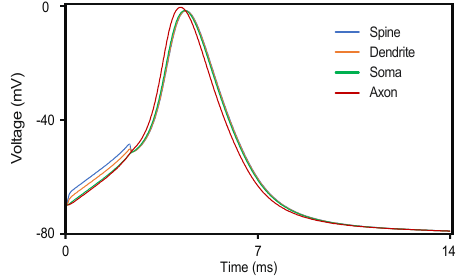}
\caption{{\bf Comparison of the action potential propagation in different neuronal regions, namely, the spine, dendritic shaft, soma and the axon.}
The value of the potential in spine is higher until the action potential is fired at the axon hillock to propagate along the axon and backpropagate into the spine.}
\label{fig:figure2Label}
\end{figure}

\begin{figure}[!h]
\centering
\includegraphics[width=1.0\linewidth]{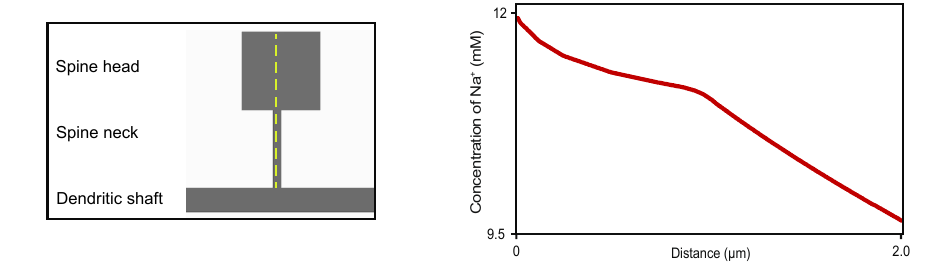}
\caption{{\bf Variation of sodium ion concentration in the spine head and neck.}
Geometry of a spine is depicted on the left. The plot on the right depicts the concentration of sodium ions along the dashed yellow line. The concentration gradient is higher in the spine neck due to its higher diffusional resistance.}
\label{fig:figure3Label}
\end{figure}

In this section we present the result of action potential propagation due to the synaptic current input at a single spine head. Fig. \ref{fig:figure2Label} compares the action potential propagation at the spine head, dendritic shaft, soma and the axon. For axon, we record the value in the region close to the axon hillock region. We observe that due to the synaptic current input at the spine, the potential is initially higher at the spine head and least at the axon until the threshold potential is attained at the axon hillock region. Once the threshold is attained, due to the action of the ion channels, the action potential forward propagates along the axon but also leads to the back-propagation in the dendritic spines. A peak action potential value of around 0 mV is numerically recorded as has been observed in \cite{gulati2023brain}. It is to be noted that the synaptic input is only present until the threshold potential is attained. Fig. \ref{fig:figure3Label} presents the concentration of sodium ions in the spine head and the spine neck along the marked yellow line at the threshold potential timeframe. We can observe that there is a significant concentration variation in the spine. Cable theory which does not consider the ionic variation is thus inaccurate for such small geometries and the Poisson-Nernst-Planck equations present a versatile framework for accurate modeling of the electrical signaling.

\subsubsection{Effect of input synaptic current density}
\begin{figure}[!h]
  \centering
\includegraphics[width=0.5\linewidth]{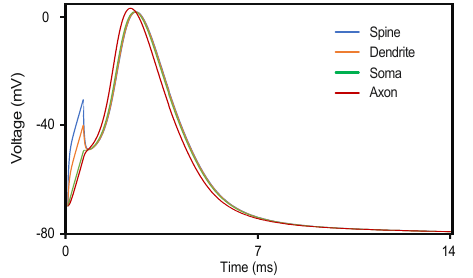}
\caption{{\bf A comparison of action potential propagation in the different neuronal regions when the input synaptic current increases by five times.}
It can be clearly observed that the potential difference between the spine and the soma region increases drastically.}
\label{fig:figure4Label}
\end{figure}

To observe the effect of spine geometry on the action potential, we start by varying the input synaptic current. On increasing the input synaptic current density by five times, we observe that the potential difference between the spine and the soma region increases drastically. This is due to the diffusional barrier offered by the spine neck. This is evident from Fig. \ref{fig:figure4Label}. The number of receptors on the spine head or the receptor density on the spine head thus plays an influential role in regulating the voltage attenuation at the soma. As expected, we also observe that the threshold potential is attained early for this case and the peak action potential is higher compared to the previous case.

\subsubsection{Effect of spine neck geometry}
\begin{figure}[!h]
  \centering
\includegraphics[width=0.5\linewidth]{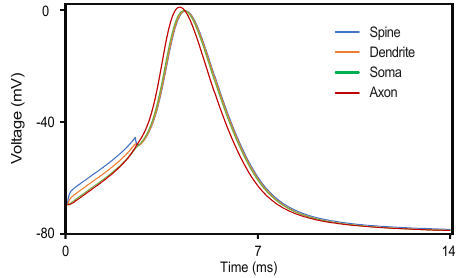}
\caption{{\bf Action potential propagation in different regions of the neuron when the spine neck length is increased.}
A peak action potential of around $0$ $mV$ is attained for this case.}
\label{fig:figure5Label}
\end{figure}

Fig. \ref{fig:figure5Label} presents the potential plots for various regions in the neuron upon increasing the spine neck length from $1 \mu m$ to $4 \mu m$. We note that the potential attenuation from the spine to the soma region is higher compared to the base case due to the higher resistance induced by the increased neck length. However, this attenuation is lesser compared to the case when the synaptic current is increased by five folds. A similar case is observed when the spine neck radius is decreased. The initial spine neck resistance is computed to be 54 $M \Omega$. 

\subsubsection{Effect of dendritic shaft geometry}
\begin{figure}[!h]
  \centering
\includegraphics[width=0.5\linewidth]{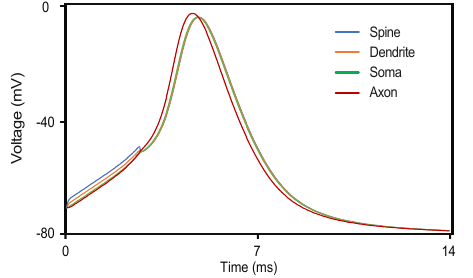}
\caption{{\bf Action potential propagation at various regions in the neuron upon increasing the width of the dendritic shaft.}
The peak action potential is lower for this case as the ion channel density needs to be higher due to larger volume of the geometry.}
\label{fig:figure6Label}
\end{figure}

\begin{figure}[!h]
  \centering
\includegraphics[width=0.5\linewidth]{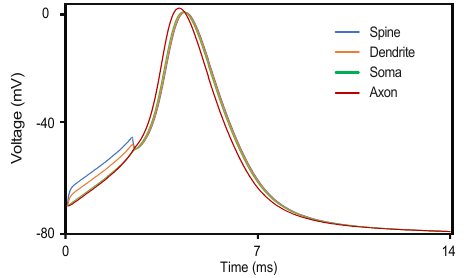}
\caption{{\bf Effect of resistance offered by the geometry of dendritic shaft.}
High diffusional resistance offered by the thin dendritic shaft as observed from the potential propagating in distinct regions in the neuron.}
\label{fig:figure7Label}
\end{figure}

After carefully studying the effect of spine geometry, we now observe the effect of the geometry of the dendritic shaft. Fig. \ref{fig:figure6Label} refers to the case when the width of the dendritic shaft is increased from $0.3 \mu m$ to $0.5 \mu m$. The potential difference between various regions of the neuron is not considerable for this case. If however the width of the dendritic shaft is reduced to $0.1$ $\mu m$, the potential difference between various regions is higher due to the increased diffusional resistance offered by the dendrite. This is demonstrated in Fig. \ref{fig:figure7Label}.

\subsection{Synaptic input at multiple spines}
\begin{figure}[!h]
 \centering
\includegraphics[width=0.5\linewidth]{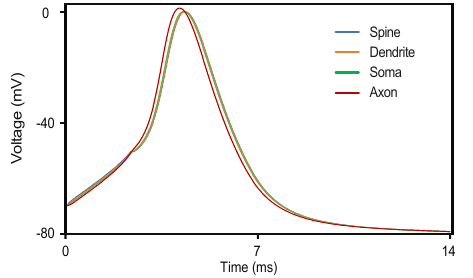}
\caption{{\bf Effect of simultaneous injection at multiple spines.}
The potential profile for action potential propagation when the synaptic current is input in multiple spines at once.}
\label{fig:figure8Label}
\end{figure}

Next, we consider the case study of the synaptic input being activated at all the four spines present in the FE geometry. As can be observed from Fig. \ref{fig:figure8Label}, we observe that before the threshold potential is attained in the axon hillock, the potential in distinct regions of the neurons is alike. After the firing of the action potential at the hillock, we observe forward propagation along the axon and the backpropagation in the spine as in the earlier cases. 

\section{Discussion}
\label{section:discussion}

\begin{figure}[!h]
  \centering
\includegraphics[width=0.5\linewidth]{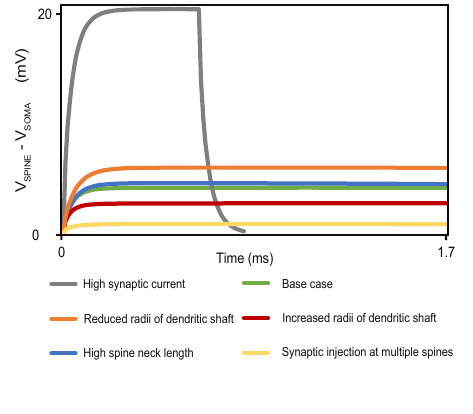}
\caption{{\bf Voltage attenuation at the soma induced by the synaptic current in the spine.}
The effect of input synaptic current density, spine geometry and the dendrite geometry on voltage attenuation at the some is evident. The spine geometry has tremendous capability to modulate the membrane potential.}
\label{fig:figure9Label}
\end{figure}

One of the key research problem of neuroscience is to get an elevated understanding of the electrical signaling in the brain. Further, in order to understand neurological diseases, it is of prime importance to understand the fundamental building blocks of the brain- the neuron. Numerical modeling has been an effective strategy to aid the experimental advances. In this work, we present a novel numerical framework of action potential propagation in a complete neuronal geometry. This model can be easily put into use / extended to understand various disease conditions, however, at this point, we focus on understanding the initiation and propagation of action potential in the neuron. We also emphasize the role of spine geometry in regulating synaptic plasticity, which has been an active area of research. The morphological changes in the spine geometry, namely the spine neck length and diameter have been known to play a key role in modulating the potential \cite{tonnesen2014nature}. The resistance of the dendritic shaft is also considerable relative to the spine. Here, we compute the initial spine neck resistance to be 54 $M \Omega$ and for the dendritic shaft to be 150 $M \Omega$ for the case of synaptic input at a single spine. Fig. \ref{fig:figure9Label} presents the voltage attenuation at the soma induced by the synaptic input at the spine. We observe that increasing the spine neck length or decreasing the radius of the spine neck leads to increasing the voltage attenuation due to the increased resistance offered by the spine neck. Thus, spine neck has a crucial role to play in synaptic plasticity. We also note that increased input synaptic current density due to the increased number of receptors being activated offers to synaptic plasticity. As reported in literature, we did not observe the effect of spine head geometry or the membrane capacitance to voltage attenuation.

The numerical framework developed here fosters our understanding of the action potential propagation through demonstrating the forward propagation along the axon as well as back propagation into the spine. We note that the voltage amplitude in the spine during forward propagation is around 23 $mV$ and during back propagation has a value of 70 $mV$. This model can certainly aid the experimental findings such as from voltage dyeing to a far better resolution. Ion channels also play a dominant role in the electrical signaling. Here, we do not study the effect of back propagating voltage attenuation at the spine based on its distance from the soma since ion channel density in the spine is not precisely known. However, once the ion channel density in the spine/ dendrite region is known, the model can be easily extended to study the effect of back propagating voltage amplitude in the spine with their distance from the soma. The ion channel density in the axon hillock region is also critical to the action potential amplitude. Note that the neuron for a rat is considered since a lot of our quantitative understanding is for a rat neuron. This model can be easily extended to other neurons such as a human neuron by incorporating the geometry, the ion channel density etc \cite{wilbers2023science}.

\section{Conclusion}

\label{section: conclusion}
Advanced experimental techniques in the last decade have provided significant insights into the spine electrophysiology. This necessitates development of a numerical framework that captures the intricate details of the neuron to an astounding resolution. Here, we extend our previous work on axonal electrophysiology using the PNP model to include the spine, the dendrite and the soma regions to observe the action potential initiation and propagation.

To the best of our knowledge, the framework presented here for spatial action potential propagation in the complete neuron geometry using the PNP model is a first-of-its-kind. Further, the role of spine geometry, dendrite geometry on the attenuation of potential from spine to the soma has been demonstrated. Both forward and backward propagation observed in the model enhance its rich capability to act as a digital twin to the neuronal experiments and provide an elevated understanding of neuronal electrophysiology.

\section{Supporting information}

\paragraph*{S1 Video.}
\label{S1_Video}
{\bf Simulation of action potential propagation in the rat neuron with myelinated axon.}  The electrical signaling in the neuron as observed from the PNP modeling. The color contours represent the voltage field. Initially the voltage is high in the spine owing to the input synaptic current induced by the binded neurotransmitter. Once the threshold is attained at the axon hillock, the forward propagation of the potential along the axon as well as the back propagation into the spines is evident from the simulation. The amplitude of the potential in spines is higher during the back propagation.


\bibliographystyle{elsarticle-num-names}
\bibliography{main}

\begin{thebibliography}{37}
\expandafter\ifx\csname natexlab\endcsname\relax\def\natexlab#1{#1}\fi
\providecommand{\url}[1]{\texttt{#1}}
\providecommand{\href}[2]{#2}
\providecommand{\path}[1]{#1}
\providecommand{\DOIprefix}{doi:}
\providecommand{\ArXivprefix}{arXiv:}
\providecommand{\URLprefix}{URL: }
\providecommand{\Pubmedprefix}{pmid:}
\providecommand{\doi}[1]{\href{http://dx.doi.org/#1}{\path{#1}}}
\providecommand{\Pubmed}[1]{\href{pmid:#1}{\path{#1}}}
\providecommand{\bibinfo}[2]{#2}
\ifx\xfnm\relax \def\xfnm[#1]{\unskip,\space#1}\fi
\bibitem[{Marion et~al.(2018)Marion, Radomski, Cramer, Galdzicki, and
  Armstrong}]{marion2018tbi}
\bibinfo{author}{C.~M. Marion}, \bibinfo{author}{K.~L. Radomski},
  \bibinfo{author}{N.~P. Cramer}, \bibinfo{author}{Z.~Galdzicki},
  \bibinfo{author}{R.~C. Armstrong},
\newblock \bibinfo{title}{Experimental traumatic brain injury identifies
  distinct early and late phase axonal conduction deficits of white matter
  pathophysiology, and reveals intervening recovery},
\newblock \bibinfo{journal}{The Journal of Neuroscience}
  \bibinfo{volume}{38(41)} (\bibinfo{year}{2018}) \bibinfo{pages}{8723--8736}.
\bibitem[{Palop et~al.(2007)Palop, Chin, Roberson, Wang, Thwin, Bien-Ly, Yoo,
  Ho, Yu, Kreitzer, Finkbeiner, Noebels, and Mucke}]{palop2007alzheimers}
\bibinfo{author}{J.~J. Palop}, \bibinfo{author}{J.~Chin},
  \bibinfo{author}{E.~D. Roberson}, \bibinfo{author}{J.~Wang},
  \bibinfo{author}{M.~T. Thwin}, \bibinfo{author}{N.~Bien-Ly},
  \bibinfo{author}{J.~Yoo}, \bibinfo{author}{K.~O. Ho}, \bibinfo{author}{G.-Q.
  Yu}, \bibinfo{author}{A.~Kreitzer}, \bibinfo{author}{S.~Finkbeiner},
  \bibinfo{author}{J.~L. Noebels}, \bibinfo{author}{L.~Mucke},
\newblock \bibinfo{title}{Aberrant excitatory neuronal activity and
  compensatory remodeling of inhibitory hippocampal circuits in mouse models of
  alzheimer's disease},
\newblock \bibinfo{journal}{Neuron} \bibinfo{volume}{55} (\bibinfo{year}{2007})
  \bibinfo{pages}{697--711}.
\bibitem[{Palop and Mucke(2010)}]{palop2010alzheimers}
\bibinfo{author}{J.~J. Palop}, \bibinfo{author}{L.~Mucke},
\newblock \bibinfo{title}{Amyloid-$\beta$ induced neuronal dysfunction in
  alzheimer’s disease: From synapses toward neural networks},
\newblock \bibinfo{journal}{Nature Neuroscience} \bibinfo{volume}{13}
  (\bibinfo{year}{2010}) \bibinfo{pages}{812--818}.
\bibitem[{Ghatak et~al.(2019)Ghatak, Dolatabadi, Trudler, Zhang, Wu, Mohata,
  Ambasudhan, Talantova, and Lipton}]{ghatak2019mechanisms}
\bibinfo{author}{S.~Ghatak}, \bibinfo{author}{N.~Dolatabadi},
  \bibinfo{author}{D.~Trudler}, \bibinfo{author}{X.~Zhang},
  \bibinfo{author}{Y.~Wu}, \bibinfo{author}{M.~Mohata},
  \bibinfo{author}{R.~Ambasudhan}, \bibinfo{author}{M.~Talantova},
  \bibinfo{author}{S.~A. Lipton},
\newblock \bibinfo{title}{Mechanisms of hyperexcitability in alzheimer’s
  disease hipsc-derived neurons and cerebral organoids vs isogenic controls},
\newblock \bibinfo{journal}{Elife} \bibinfo{volume}{8} (\bibinfo{year}{2019})
  \bibinfo{pages}{e50333}.
\bibitem[{Chaudhury et~al.(2015)Chaudhury, Liu, and
  Han}]{chaudhury2015depression}
\bibinfo{author}{D.~Chaudhury}, \bibinfo{author}{H.~Liu},
  \bibinfo{author}{M.-H. Han},
\newblock \bibinfo{title}{Neuronal correlates of depression},
\newblock \bibinfo{journal}{Cellular and molecular life sciences}
  \bibinfo{volume}{72(24)} (\bibinfo{year}{2015}) \bibinfo{pages}{4825--4848}.
\bibitem[{Qian and Sejnowski(1989)}]{qian1999pnp}
\bibinfo{author}{N.~Qian}, \bibinfo{author}{T.~Sejnowski},
\newblock \bibinfo{title}{An electro-diffusion model for computing membrane
  potentials and ionic concentrations in branching dendrites, spines and
  axons},
\newblock \bibinfo{journal}{Biological Cybernetics} \bibinfo{volume}{62}
  (\bibinfo{year}{1989}) \bibinfo{pages}{1--15}.
\bibitem[{Lagache et~al.(2019)Lagache, Jayant, and
  Yuste}]{lagache2019compneuroscience}
\bibinfo{author}{T.~Lagache}, \bibinfo{author}{K.~Jayant},
  \bibinfo{author}{R.~Yuste},
\newblock \bibinfo{title}{Electrodiffusion models of synaptic potentials in
  dendritic spines},
\newblock \bibinfo{journal}{Journal of Computational Neuroscience}
  \bibinfo{volume}{47} (\bibinfo{year}{2019}) \bibinfo{pages}{77--89}.
\bibitem[{Cartailler and Holcman(2018)}]{jerome2018physical}
\bibinfo{author}{J.~Cartailler}, \bibinfo{author}{D.~Holcman},
\newblock \bibinfo{title}{Electrical transient laws in neuronal microdomains
  based on electro-diffusion},
\newblock \bibinfo{journal}{Physical Chemistry Chemical Physics}
  \bibinfo{volume}{20(32)} (\bibinfo{year}{2018})
  \bibinfo{pages}{21062--21067}.
\bibitem[{Cartailler et~al.(2018)Cartailler, Kwon, Yuste, and
  Holcman}]{cartailler2018neuron}
\bibinfo{author}{J.~Cartailler}, \bibinfo{author}{T.~Kwon},
  \bibinfo{author}{R.~Yuste}, \bibinfo{author}{D.~Holcman},
\newblock \bibinfo{title}{Deconvolution of voltage sensor time series and
  electro-diffusion modeling reveal the role of spine geometry in controlling
  synaptic strength},
\newblock \bibinfo{journal}{Neuron} \bibinfo{volume}{97(5)}
  (\bibinfo{year}{2018}) \bibinfo{pages}{1126--1136}.
\bibitem[{Palmer and Stuart(2009)}]{palmer2009neuroscience}
\bibinfo{author}{L.~M. Palmer}, \bibinfo{author}{G.~J. Stuart},
\newblock \bibinfo{title}{Membrane potential changes in dendritic spines during
  action potentials and synaptic input},
\newblock \bibinfo{journal}{The Journal of Neuroscience}
  \bibinfo{volume}{29(21)} (\bibinfo{year}{2009}) \bibinfo{pages}{6897--6903}.
\bibitem[{Holcman and Yuste(2015)}]{holcman2015nature}
\bibinfo{author}{D.~Holcman}, \bibinfo{author}{R.~Yuste},
\newblock \bibinfo{title}{The new nanophysiology: regulation of ionic flow in
  neuronal subcompartments},
\newblock \bibinfo{journal}{Nature Reviews Neuroscience}
  \bibinfo{volume}{16(11)} (\bibinfo{year}{2015}) \bibinfo{pages}{685--692}.
\bibitem[{Acker et~al.(2016)Acker, Hoyos, and Loew}]{acker2016eneuro}
\bibinfo{author}{C.~D. Acker}, \bibinfo{author}{E.~Hoyos},
  \bibinfo{author}{L.~M. Loew},
\newblock \bibinfo{title}{Epsps measured in proximal dendritic spines of
  cortical pyramidal neurons},
\newblock \bibinfo{journal}{eNeuro} \bibinfo{volume}{3(2)}
  (\bibinfo{year}{2016}) \bibinfo{pages}{1--13}.
\bibitem[{Kwon et~al.(2017)Kwon, Sakamoto, Peterka, and Yuste}]{kwon2017cell}
\bibinfo{author}{T.~Kwon}, \bibinfo{author}{M.~Sakamoto},
  \bibinfo{author}{D.~S. Peterka}, \bibinfo{author}{R.~Yuste},
\newblock \bibinfo{title}{Attenuation of synaptic potentials in dendritic
  spines},
\newblock \bibinfo{journal}{Cell Reports} \bibinfo{volume}{20(5)}
  (\bibinfo{year}{2017}) \bibinfo{pages}{1100--1110}.
\bibitem[{Jayant et~al.(2017)Jayant, Hirtz, Plante, Tsai, De~Boer, Semonche,
  Peterka, Owen, Sahin, Shepard, and Yuste}]{jayant2017nature}
\bibinfo{author}{K.~Jayant}, \bibinfo{author}{J.~J. Hirtz},
  \bibinfo{author}{I.~J.-L. Plante}, \bibinfo{author}{D.~M. Tsai},
  \bibinfo{author}{W.~D. De~Boer}, \bibinfo{author}{A.~Semonche},
  \bibinfo{author}{D.~S. Peterka}, \bibinfo{author}{J.~S. Owen},
  \bibinfo{author}{O.~Sahin}, \bibinfo{author}{K.~L. Shepard},
  \bibinfo{author}{R.~Yuste},
\newblock \bibinfo{title}{Targeted intracellular voltage recordings from
  dendritic spines using quantum-dot-coated nanopipettes},
\newblock \bibinfo{journal}{Nature nanotechnology} \bibinfo{volume}{12(4)}
  (\bibinfo{year}{2017}) \bibinfo{pages}{335--342}.
\bibitem[{Yuste(2013)}]{yuste2013neuroscience}
\bibinfo{author}{R.~Yuste},
\newblock \bibinfo{title}{Electrical compartmentalization in dendritic spines},
\newblock \bibinfo{journal}{Annual review of neuroscience} \bibinfo{volume}{36}
  (\bibinfo{year}{2013}) \bibinfo{pages}{429--449}.
\bibitem[{Harnett et~al.(2012)Harnett, Makara, Spruston, Kath, and
  Magee}]{harnett2012nature}
\bibinfo{author}{M.~T. Harnett}, \bibinfo{author}{J.~K. Makara},
  \bibinfo{author}{N.~Spruston}, \bibinfo{author}{W.~L. Kath},
  \bibinfo{author}{J.~C. Magee},
\newblock \bibinfo{title}{Synaptic amplification by dendritic spines enhances
  input cooperativity},
\newblock \bibinfo{journal}{Nature} \bibinfo{volume}{491(7425)}
  (\bibinfo{year}{2012}) \bibinfo{pages}{599--602}.
\bibitem[{Araya et~al.(2006)Araya, Jiang, Eisenthal, and Yuste}]{araya2006pnas}
\bibinfo{author}{R.~Araya}, \bibinfo{author}{J.~Jiang}, \bibinfo{author}{K.~B.
  Eisenthal}, \bibinfo{author}{R.~Yuste},
\newblock \bibinfo{title}{The spine neck filters membrane potentials},
\newblock \bibinfo{journal}{Proceedings of the National Academy of Sciences}
  \bibinfo{volume}{103(47)} (\bibinfo{year}{2006})
  \bibinfo{pages}{17961--17966}.
\bibitem[{Hoffman et~al.(1997)Hoffman, Magee, Colbert, and
  Johnston}]{hoffman1997nature}
\bibinfo{author}{D.~A. Hoffman}, \bibinfo{author}{J.~C. Magee},
  \bibinfo{author}{C.~M. Colbert}, \bibinfo{author}{D.~Johnston},
\newblock \bibinfo{title}{K+ channel regulation of signal propagation in
  dendrites of hippocampal pyramidal neurons},
\newblock \bibinfo{journal}{Nature} \bibinfo{volume}{387(6636)}
  (\bibinfo{year}{1997}) \bibinfo{pages}{869--875}.
\bibitem[{Colbert and Johnston(1996)}]{colbert1996neuroscience}
\bibinfo{author}{C.~M. Colbert}, \bibinfo{author}{D.~Johnston},
\newblock \bibinfo{title}{Axonal action-potential initiation and na+ channel
  densities in the soma and axon initial segment of subicular pyramidal
  neurons},
\newblock \bibinfo{journal}{Journal of Neuroscience} \bibinfo{volume}{16(21)}
  (\bibinfo{year}{1996}) \bibinfo{pages}{6676--6686}.
\bibitem[{Hodgkin and Huxley(1952)}]{hodgkin1952huxley}
\bibinfo{author}{A.~Hodgkin}, \bibinfo{author}{A.~Huxley},
\newblock \bibinfo{title}{A quantitative description of membrane current and
  its application to conduction and excitation in nerve},
\newblock \bibinfo{journal}{The Journal of Physiology} \bibinfo{volume}{117(4)}
  (\bibinfo{year}{1952}) \bibinfo{pages}{500--544}.
\bibitem[{Huxley(1959)}]{huxley1959resistance}
\bibinfo{author}{A.~F. Huxley},
\newblock \bibinfo{title}{Ion movements during nerve activity},
\newblock \bibinfo{journal}{Second conference on physicochemical mechanism of
  nerve activity and second conference on muscular contraction}
  \bibinfo{volume}{81(2)} (\bibinfo{year}{1959}) \bibinfo{pages}{221--246}.
\bibitem[{Cohen et~al.(2020)Cohen, Popovic, Klooster, Weil, M\"{o}bius, Nave,
  and Kole}]{cohen2020cell}
\bibinfo{author}{C.~C. Cohen}, \bibinfo{author}{M.~A. Popovic},
  \bibinfo{author}{J.~Klooster}, \bibinfo{author}{M.-T. Weil},
  \bibinfo{author}{W.~M\"{o}bius}, \bibinfo{author}{K.-A. Nave},
  \bibinfo{author}{M.~H. Kole},
\newblock \bibinfo{title}{Saltatory conduction along myelinated axons involves
  a periaxonal nanocircuit},
\newblock \bibinfo{journal}{Cell} \bibinfo{volume}{180} (\bibinfo{year}{2020})
  \bibinfo{pages}{311--322}.
\bibitem[{Pods et~al.(2013)Pods, Schonke, and Bastian}]{pods2013pnp}
\bibinfo{author}{J.~Pods}, \bibinfo{author}{J.~Schonke},
  \bibinfo{author}{P.~Bastian},
\newblock \bibinfo{title}{Electrodiffusion models of neurons and extracellular
  space using the poisson-nernst-planck equations—numerical simulation of the
  intra- and extracellular potential for an axon model},
\newblock \bibinfo{journal}{Biophysical Journal} \bibinfo{volume}{105}
  (\bibinfo{year}{2013}) \bibinfo{pages}{242--254}.
\bibitem[{Lopreore et~al.(2008)Lopreore, Bartol, Coggan, Keller, Sosinsky,
  Ellisman, and Sejnowski}]{lopreore2008pnp2}
\bibinfo{author}{C.~L. Lopreore}, \bibinfo{author}{T.~M. Bartol},
  \bibinfo{author}{J.~S. Coggan}, \bibinfo{author}{D.~X. Keller},
  \bibinfo{author}{G.~E. Sosinsky}, \bibinfo{author}{M.~H. Ellisman},
  \bibinfo{author}{T.~J. Sejnowski},
\newblock \bibinfo{title}{Computational modeling of three-dimensional
  electrodiffusion in biological systems: Application to the node of ranvier},
\newblock \bibinfo{journal}{Biophysical Journal} \bibinfo{volume}{95}
  (\bibinfo{year}{2008}) \bibinfo{pages}{2624--2635}.
\bibitem[{Dione et~al.(2016)Dione, Deteix, Briffard, and
  Chamberland}]{dione2016pnp1}
\bibinfo{author}{I.~Dione}, \bibinfo{author}{J.~Deteix},
  \bibinfo{author}{T.~Briffard}, \bibinfo{author}{E.~Chamberland},
\newblock \bibinfo{title}{Improved simulation of electrodiffusion in the node
  of ranvier by mesh adaptation},
\newblock \bibinfo{journal}{PLOS ONE} \bibinfo{volume}{11(8)}
  (\bibinfo{year}{2016}) \bibinfo{pages}{e0161318}.
\bibitem[{Gulati and Rudraraju(2023)}]{gulati2023brain}
\bibinfo{author}{R.~Gulati}, \bibinfo{author}{S.~Rudraraju},
\newblock \bibinfo{title}{Spatio-temporal modeling of saltatory conduction in
  neurons using poisson–nernst–planck treatment and estimation of
  conduction velocity},
\newblock \bibinfo{journal}{Brain Multiphysics} \bibinfo{volume}{4}
  (\bibinfo{year}{2023}) \bibinfo{pages}{100061--100072}.
\bibitem[{Popovic et~al.(2015)Popovic, Carnevale, Rozsa, and
  Zecevic}]{popovic2015nature}
\bibinfo{author}{M.~A. Popovic}, \bibinfo{author}{N.~Carnevale},
  \bibinfo{author}{B.~Rozsa}, \bibinfo{author}{D.~Zecevic},
\newblock \bibinfo{title}{Electrical behaviour of dendritic spines as revealed
  by voltage imaging},
\newblock \bibinfo{journal}{Nature communications} \bibinfo{volume}{6(1)}
  (\bibinfo{year}{2015}) \bibinfo{pages}{8436--8448}.
\bibitem[{Auddya et~al.(2021)Auddya, Zhang, Gulati, Vasan, Garikipati,
  Rangamani, and Rudraraju}]{deb2021shellMechanics}
\bibinfo{author}{D.~Auddya}, \bibinfo{author}{X.~Zhang},
  \bibinfo{author}{R.~Gulati}, \bibinfo{author}{R.~Vasan},
  \bibinfo{author}{K.~Garikipati}, \bibinfo{author}{P.~Rangamani},
  \bibinfo{author}{S.~Rudraraju},
\newblock \bibinfo{title}{Biomembranes undergo complex, non-axisymmetric
  deformations governed by kirchhoff-love kinematics and revealed by a three
  dimensional computational framework},
\newblock \bibinfo{journal}{Proceedings of the Royal Society A}
  \bibinfo{volume}{477(2255)} (\bibinfo{year}{2021}) \bibinfo{pages}{20210246}.
\bibitem[{Bar-Kochba et~al.(2016)Bar-Kochba, Scimone, Estrada, and
  Franck}]{bar2016strain}
\bibinfo{author}{E.~Bar-Kochba}, \bibinfo{author}{M.~T. Scimone},
  \bibinfo{author}{J.~B. Estrada}, \bibinfo{author}{C.~Franck},
\newblock \bibinfo{title}{Strain and rate-dependent neuronal injury in a 3d in
  vitro compression model of traumatic brain injury},
\newblock \bibinfo{journal}{Scientific Reports} \bibinfo{volume}{6}
  (\bibinfo{year}{2016}) \bibinfo{pages}{1--11}.
\bibitem[{Li et~al.(2011)Li, Fan, Ji, Qiu, and Li}]{li2011effects}
\bibinfo{author}{W.~Li}, \bibinfo{author}{Q.~Fan}, \bibinfo{author}{Z.~Ji},
  \bibinfo{author}{X.~Qiu}, \bibinfo{author}{Z.~Li},
\newblock \bibinfo{title}{The effects of irreversible electroporation (ire) on
  nerves},
\newblock \bibinfo{journal}{PloS one} \bibinfo{volume}{6}
  (\bibinfo{year}{2011}) \bibinfo{pages}{e18831}.
\bibitem[{Estrada et~al.(2021)Estrada, Cramer~III, Scimone, Buyukozturk, and
  Franck}]{estrada2021neural}
\bibinfo{author}{J.~B. Estrada}, \bibinfo{author}{H.~C. Cramer~III},
  \bibinfo{author}{M.~T. Scimone}, \bibinfo{author}{S.~Buyukozturk},
  \bibinfo{author}{C.~Franck},
\newblock \bibinfo{title}{Neural cell injury pathology due to high-rate
  mechanical loading},
\newblock \bibinfo{journal}{Brain Multiphysics} \bibinfo{volume}{2}
  (\bibinfo{year}{2021}) \bibinfo{pages}{100034}.
\bibitem[{Palop and Mucke(2016)}]{palop2016network}
\bibinfo{author}{J.~J. Palop}, \bibinfo{author}{L.~Mucke},
\newblock \bibinfo{title}{Network abnormalities and interneuron dysfunction in
  alzheimer disease},
\newblock \bibinfo{journal}{Nature Reviews Neuroscience} \bibinfo{volume}{17}
  (\bibinfo{year}{2016}) \bibinfo{pages}{777--792}.
\bibitem[{Kim et~al.(2007)Kim, Carey, Wang, Ingano, Binshtok, Wertz,
  Pettingell, He, Lee, Woolf et~al.}]{kim2007bace1}
\bibinfo{author}{D.~Y. Kim}, \bibinfo{author}{B.~W. Carey},
  \bibinfo{author}{H.~Wang}, \bibinfo{author}{L.~A. Ingano},
  \bibinfo{author}{A.~M. Binshtok}, \bibinfo{author}{M.~H. Wertz},
  \bibinfo{author}{W.~H. Pettingell}, \bibinfo{author}{P.~He},
  \bibinfo{author}{V.~M.-Y. Lee}, \bibinfo{author}{C.~J. Woolf}, et~al.,
\newblock \bibinfo{title}{Bace1 regulates voltage-gated sodium channels and
  neuronal activity},
\newblock \bibinfo{journal}{Nature cell biology} \bibinfo{volume}{9}
  (\bibinfo{year}{2007}) \bibinfo{pages}{755--764}.
\bibitem[{Freidberg(2008)}]{pnp2008gauss2pnp}
\bibinfo{author}{J.~P. Freidberg}, \bibinfo{title}{Plasma Physics and Fusion
  Energy}, volume~\bibinfo{volume}{1}, \bibinfo{publisher}{Cambridge University
  Press}, \bibinfo{year}{2008}.
\bibitem[{Git(2022)}]{GitRepo2022}
\bibinfo{title}{Code repository for action potential propagation along the
  neuronal axon and spatio-temporal saltatory conduction},
  \bibinfo{howpublished}{https://github.com/cmmg/neuronalActionPotential},
  \bibinfo{year}{2022}.
\bibitem[{T{\o}nnesen et~al.(2014)T{\o}nnesen, Katona, R{\'o}zsa, and
  N{\"a}gerl}]{tonnesen2014nature}
\bibinfo{author}{J.~T{\o}nnesen}, \bibinfo{author}{G.~Katona},
  \bibinfo{author}{B.~R{\'o}zsa}, \bibinfo{author}{U.~V. N{\"a}gerl},
\newblock \bibinfo{title}{Spine neck plasticity regulates compartmentalization
  of synapses},
\newblock \bibinfo{journal}{Nature neuroscience} \bibinfo{volume}{17}
  (\bibinfo{year}{2014}) \bibinfo{pages}{678--685}.
\bibitem[{Wilbers et~al.(2023)Wilbers, Metodieva, Duverdin, Heyer, Galakhova,
  Mertens, Versluis, Baayen, Idema, Noske, and Verburg}]{wilbers2023science}
\bibinfo{author}{R.~Wilbers}, \bibinfo{author}{V.~Metodieva},
  \bibinfo{author}{S.~Duverdin}, \bibinfo{author}{D.~Heyer},
  \bibinfo{author}{A.~Galakhova}, \bibinfo{author}{E.~Mertens},
  \bibinfo{author}{T.~Versluis}, \bibinfo{author}{J.~Baayen},
  \bibinfo{author}{S.~Idema}, \bibinfo{author}{D.~Noske},
  \bibinfo{author}{N.~Verburg},
\newblock \bibinfo{title}{Human voltage-gated na+ and k+ channel properties
  underlie sustained fast ap signaling},
\newblock \bibinfo{journal}{Science Advances} \bibinfo{volume}{9}
  (\bibinfo{year}{2023}) \bibinfo{pages}{eade3300}.

\end{thebibliography}

\end{document}